\documentclass[aps,prx,twocolumn,english,balance,superscriptaddress,floats,showpacs,prb,footinbib]{revtex4-2}
\usepackage[latin9]{inputenc}
\usepackage{amsmath}
\usepackage{amssymb}
\usepackage{graphicx}
\usepackage{esint}
\usepackage{subfigure}
\usepackage{multirow}
\usepackage{mathtools}
\usepackage{xcolor}
\makeatletter

\newcommand{\beq}{\begin{equation}}
	\newcommand{\eeq}{\end{equation}}
\newcommand{\bea}{\begin{eqnarray}}
	\newcommand{\eea}{\end{eqnarray}}
\newcommand{\bwt}{\begin{widetext}}
	\newcommand{\ewt}{\end{widetext}}
\@ifundefined{textcolor}{}
{%
	\definecolor{BLACK}{gray}{0}
	\definecolor{WHITE}{gray}{1}
	\definecolor{RED}{rgb}{1,0,0}
	\definecolor{GREEN}{rgb}{0,1,0}
	\definecolor{BLUE}{rgb}{0,0,1}
	\definecolor{CYAN}{cmyk}{1,0,0,0}
	\definecolor{MAGENTA}{cmyk}{0,1,0,0}
	\definecolor{YELLOW}{cmyk}{0,0,1,0}
}

\newcommand{\br}{\mathbf{r}}

\newcommand{\bK}{\mathbf{K}}

\newcommand{\bX}{\mathbf{X}}
\newcommand{\bG}{\mathbf{G}}

\newcommand{\fvec}[1]{\boldsymbol{#1}}

\newcommand{\half}{\frac{1}{2}}

\newcommand{\rmd}{{\rm d}}

\makeatother

\begin{document}
	
	\title{Continuum effective Hamiltonian for graphene bilayers for an arbitrary smooth lattice deformation from microscopic theories}

	\author{Oskar Vafek}
	\email{vafek@magnet.fsu.edu}
	\affiliation{National High Magnetic Field Laboratory, Tallahassee, Florida, 32310, USA}
	\affiliation{Department of Physics, Florida State University, Tallahassee, Florida 32306, USA}
	
		\author{Jian Kang}
	\email{kangjian@shanghaitech.edu.cn}
	\affiliation{School of Physical Science and Technology, ShanghaiTech University, Shanghai 200031, China}
	
	\begin{abstract}
		We provide a systematic real space derivation of the continuum Hamiltonian for a graphene bilayer starting from a microscopic lattice theory, allowing for an arbitrary inhomogeneous smooth lattice deformation, including a twist. Two different microscopic models are analyzed: first, a Slater-Koster like model and second, ab-initio derived model. 
		We envision that our effective Hamiltonian can be used in conjunction with an experimentally determined atomic lattice deformation in twisted bilayer graphene in a specific device to predict and compare the electronic spectra with scanning tunneling spectroscopy measurements.
		As a byproduct, our approach provides electron-phonon couplings in the continuum Hamiltonian from microscopic models for any bilayer stacking. In the companion paper we analyze in detail the continuum models for relaxed atomic configurations of magic angle twisted bilayer graphene.
	\end{abstract}
	
	\maketitle
	\section{Introduction}
	Observation of the correlated electron phenomena~\cite{Pablo1}, including superconductivity~\cite{Pablo2}, in the vicinity of the first magic angle in twisted bilayer graphene~\cite{NetoPRL07,MagaudNL10,Barticevic,BMModel} led to a large number of experimental~\cite{Cory1,David,Young,Dmitry1,Yazdani,Ashoori,Eva,Stevan,Young2,Dmitry2,Yazdani2,Shahal,Abhay19,Stevan19,Young3,YuanCao2021,Shahal2,Yacoby,Yacoby2,PabloNature2021,Young3,Zeldov,Young4,YoungCDW,JiaSC,JiaSOC,YazdaniSC} and theoretical studies~\cite{Senthil1,LiangPRX1,KangVafekPRX,FengchengSC,GuineaPNAS,Balents19,BJYangPRX,Bernevig1,Leon2,Dai1,Grisha,KangVafekPRL,KangVafekPRB,Senthil2,Dai2,MacDonald,Zaletel1,Zaletel2,ZaletelDMRG,NickKekule,BernevigTBG}  of this remarkable physical system.
	Although the main experimental findings~\cite{Pablo1,Pablo2,Cory1,David,Young} were reproduced by a number of experimental groups, there is a nagging lack of reproducibility in the finer details of the physical characteristics of devices, even when manufactured within a same lab and even within a same device.
	This is likely due to spatial inhomogeneity in the twist angle~\cite{Zeldov,Shahal,Young2,Yazdani2} 
	and unintentional strain~\cite{Yazdani} produced during the device fabrication, or more generally, due to lattice deformations which vary over distances long compared to the microscopic spacing between neighboring carbon atoms.
	
	It is thus being recognized that the twist angle is not the only parameter controlling the physics of a specific device~\cite{Yacoby}. This fact motivates a development of a theory whose input would be more than just the twist angle $\theta$, Fermi velocity $v_F$ and the two inter-layer tunneling constants through the AA ($w_0$) and AB ($w_1$) regions, as is the case for (the slight generalization of) the original Bistritzer-MacDonald (BM) model, but instead, the input would be a smooth and possibly inhomogeneous configuration of the atomic displacement field. This configuration could in principle be extracted from topography measured using a scanning tunneling microscope~\cite{Eva,Yazdani,Abhay19,Stevan19,AllanPRR21} or from Bragg interferometry\cite{BediakoNatMat2021}.
	
	The goal of this paper is to provide a systematic derivation of such a continuum Hamiltonian for an arbitrary smooth atomic displacement $\fvec u_j(\fvec r)$ starting from a microscopic {\it ab initio} calibrated tight-binding model on the carbon lattice. Expanding in gradients of $\fvec u_j(\fvec r)$ and of slowly varying envelope of the graphene's $\bK$ and $\bK'$ Bloch functions one can achieve any desired accuracy when comparing with the microscopic model, as we demonstrate in the companion paper for the relaxed atomic configurations of the magic angle twisted bilayer graphene. Here, we provide the general formulas for two different microscopic models. For the first, we consider a microscopic hopping function which depends only on the separation between two carbon atoms, as is the case in the Slater-Koster type models~\cite{Ando2001,Uryu2004,SlaterKoster1954,MagaudNL10,KoshinoPRB12,KangVafekPRX}. For the second, we allow for dependence of the inter-layer tunneling terms on the relative orientation of the interatomic separation vector and the nearest neighbor bonds, as is the case in the microscopic model derived from density functional theory (DFT) determined Wannier states of the monolayer (and untwisted bilayer) graphene's conduction and valance bands in Ref.~\cite{KaxirasPRB16}, as well as for the configuration dependence of the on-site term.
	
	The method which we develop here is inspired by the approach advanced by Balents~\cite{Balents19}, but strives to go beyond it in several ways. 
	
	First, the continuum Hamiltonian is derived entirely from the microscopic tight binding models. As a consequence, all the parameters in the continuum Hamiltonian can be expressed by suitable moments of the hopping functions and the lattice distortions, yielding realistic values of the electron-phonon couplings as a byproduct. This allows for a direct comparison of the theory with the experiments if the deformed positions of atoms are measured by local probes. 
	
	Second, when applied to the twisted bilayer graphene, the continuum model derived here goes beyond the BM model\cite{BMModel,Balents19} by systematically including higher order gradient terms i.e.~gradients of both the slowly varying envelope of the fermion fields and of the atomic displacement fields. Because they are derived directly in real space via a gradient expansion, each term in our continuum theory is local~\cite{MacDonalNonlocal} (although, of course, not necessarily just `contact').
	In contrast, the continuum models in Refs.~\cite{BMModel,KoshinoPRB20} are obtained in the momentum space which makes any treatment of spatial inhomogeneity inconvenient, and when translated to real space, the existing models include only (some of the) first order gradient terms.
	The motivation for including higher order gradient terms is directly related to the physics of the magic angle. For a twist angle $\theta$ near the magic value $1.1^{\circ}$, the estimate of the energy scale of the leading order terms constituting the BM model\cite{BMModel,Balents19} can be obtained by multiplying the Fermi velocity $v_F$ and the typical momentum deviation from the Dirac point, $\hbar v_F|\bK|\theta\sim 200$meV for the intra-layer term, and $\sim 100$meV for the contact inter-layer term. The second order intra-layer derivative terms and the first order derivative in fermion and atomic displacement fields inter-layer terms are smaller by the factor of $\sim|\fvec K| a \theta = 4\pi\theta/3\sim 0.08$, seemingly justifying their omission (for definitions of various parameters mentioned, see the next section). As is well known, however, at the magic angle the non-interacting bandwidth is anomalously smaller than the scale of the leading order terms by at least an order of magnitude, making the higher order terms comparable to the non-interacting narrow bandwidth~\cite{GuineaModel}. Moreover, even if smaller, they can be of similar order to the scale of Coulomb interaction, and they break particle-hole symmetry~\cite{Bernevig1,Leon2,Dai1,KaxirasGradient}, thus lifting degeneracy of the ground state manifold in strong coupling~\cite{KangVafekPRL,Zaletel2,BernevigTBG,NickKekule,VafekKangPRL20}. Therefore, it is desirable to study their effects systematically as we do here and the companion paper.
	
	Finally, it was recognized\cite{LiangPRX1} that atomic relaxation of twisted bilayer graphene near the first magic angle leads to an increase in the size of the AB stacked regions of the moire pattern at the expense of the AA stacked regions, as compared to the structure resulting from a simple rigid twist. With few exceptions~\cite{KoshinoPRB20}, such relaxation has been modeled as a simple change of AA and AB tunneling parameters $w_0$ and $w_1$ respectively. However, because the difference between $w_0$ and $w_1$ arises from lattice distortions, such relaxation must include pseudo-magnetic vector potential terms --given by combinations of {\em first order} spatial derivatives of the atomic displacement fields-- in the intra-layer Hamiltonian. Within the gradient expansion, such terms appear at the same order as the intra-layer first order gradient of the slow Fermi fields i.e. same order as the massless Dirac terms. Indeed, we find that such terms are comparable to the inter-layer tunneling terms $w_{0,1}$ included in the BM model, and therefore there is no a'priori justification for neglecting them.

\begin{widetext}
		\begin{figure*}[htb]
			\centering
			\includegraphics[width=1.9\columnwidth]{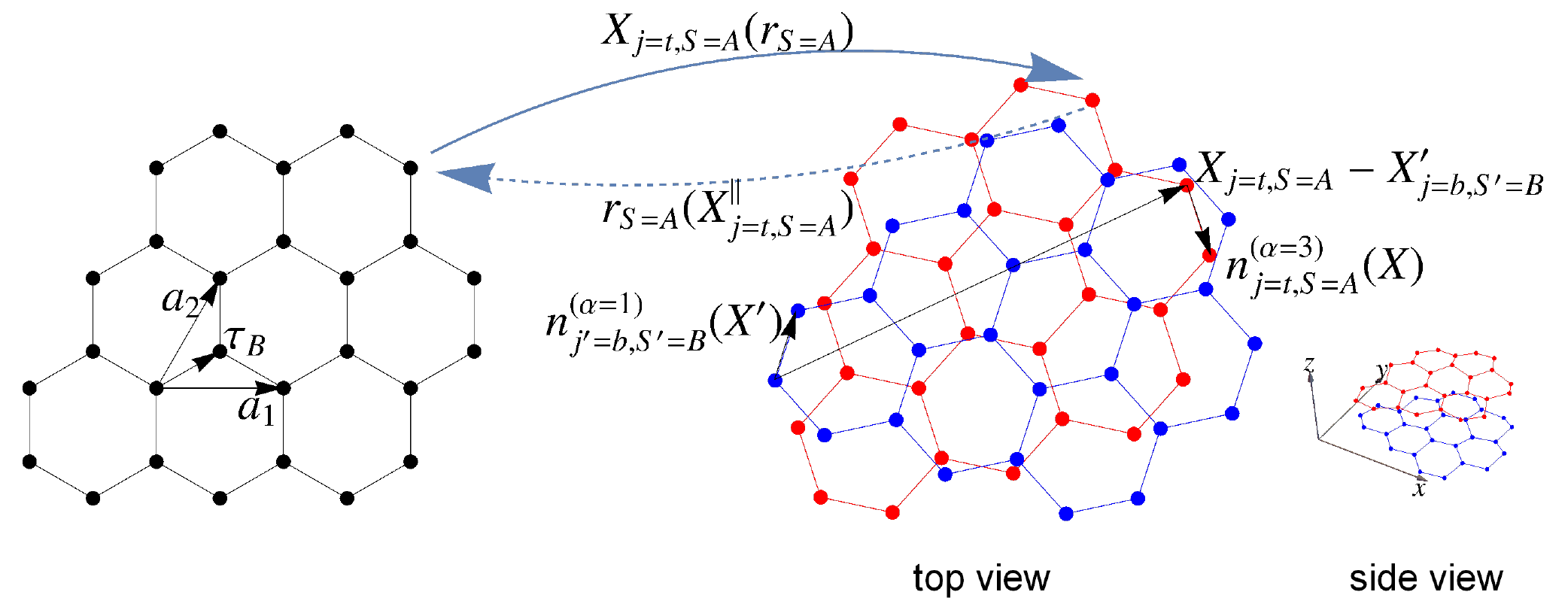}
			\caption{Schematic illustration of the one-to-one mapping between (left) the undistorted atomic position $\fvec r_S$ and (right, top view) the distorted atomic position $\fvec X_{j, S}$, where red is for the top layer and blue for the bottom layer. The separation between two carbon atoms $\fvec X_{j,S} - \fvec X'_{j',S'}$ and the corresponding nearest neighbor vector $\fvec n_{j,S}$ and $\fvec n_{j',S'}$ are labeled by black arrows. The distortion described by $\fvec X_{j, S}$ also includes the lattice corrugation, as shown in the side view.}
			\label{Fig:Schematic}
		\end{figure*}
	\end{widetext}
	
		\section{Microscopic derivation of the continuum low energy model for an arbitrary smooth lattice deformation} \label{Sec:DeriveEffModel}
	
	In order to derive the effective continuum Hamiltonian  from the microscopic tight binding model, we start by noting that generally, the distorted position $\fvec X_{j,S}$ of a carbon atom in the layer $j$ and sublattice $S$ can be expressed as
	\begin{align}
		\fvec X_{j,S} & =  \fvec r_S + \fvec u_{j, S}(\fvec r_S)\equiv \fvec X_{j,S}(\fvec r_S), \label{Eqn:XPos} \\
		\fvec u_{j, S}(\fvec r_S) & = \fvec u^{\parallel}_{j,S}\left(\br_{S}\right) + \fvec u^{\perp}_{j,S}(\fvec \br_S),\label{Eqn:udecomp}
	\end{align}
	where $\fvec r_S = n_1 \fvec a_1 + n_2 \fvec a_2 + \fvec \tau_{S}$ is the reference undistorted position of the carbon atom within a honeycomb lattice (see Fig.~\ref{Fig:Schematic}), with $n_{1,2}$ being integers. The basis vectors of the undistorted lattice are $\fvec \tau_A = 0$ for the sublattice A and $\fvec \tau_B = (\fvec a_1 + \fvec a_2)/3$ for sublattice B, where $\fvec a_1 = a(1, 0)$ and $\fvec a_2 = a(\frac12,\frac{\sqrt{3}}2)$ are the two primitive lattice vectors and $a=0.246$nm being the lattice constant.
	The displacement $\fvec u_{j, S}(\fvec r_S)$ describes the deviation from the undistorted position of the carbon atoms. It is general enough to account for twist, in-plane relaxation, out-of-plane corrugation and strain, as well as any possible difference between atomic displacements of the two sublattices. The vector $\fvec u_{j, S}(\fvec r_S)$ in Eq.(\ref{Eqn:udecomp}) is decomposed into an in-plane component $\fvec u^{\parallel}_{j,S}\left(\br_{S}\right)$ and an out-of-plane component $\fvec u^{\perp}_{j,S}(\fvec \br_S)$. Its explicit dependence on the undistorted lattice point $\fvec r_S$ is referred to as the Lagrangian coordinates~\cite{Balents19,ChaikinLubensky}.
	
	Although we start with the Lagrangian formulation, we will reach a point in our derivation where we switch to the more convenient Eulerian coordinates~\cite{Balents19,ChaikinLubensky}, where the displacements are expressed in terms of the actual in-plane position of the atoms $\fvec X^\parallel_{j,S}$ as opposed to the undistorted positions $\fvec r_S$. Because each monolayer graphene sheet is assumed not to fold, there is a one-to-one
	mapping between $\fvec r_S$ and $\fvec X^\parallel_{j,S}$. If it folded, there would be overhangs for a sheet, and two different positions $\fvec r_S$ would
	map onto the same $\fvec X^\parallel_{j,S}$. Without overhangs, we can therefore adopt the Monge ``gauge''\cite{ChaikinLubensky} and use the Eulerian coordinates and write
	\begin{align}\label{Eqn:EulerCoord}
		\fvec X_{j,S} & = \fvec r_S + \fvec U^{\parallel}_{j,S}(\fvec X^{\parallel}_{j,S}) + \fvec U^{\perp}_{j,S}(\fvec X^\parallel_{j,S}).
	\end{align}
	The displacement functions $\fvec U^{\parallel,\perp}_{j,S}$ now depend on the actual in-plane location of the distorted atoms which can be determined by solving Eq.\ref{Eqn:XPos} for $\fvec{r}_S$ in terms of $\fvec X^\parallel_{j,S}$ and then expressing the displacement fields in terms of $\fvec X^\parallel_{j,S}$.
	
	To illustrate the main idea, in this section we allow the hopping amplitude $t$ to depend only on the separation of the two carbon atoms $\fvec X_{j,S} - \fvec X'_{j', S'}$. In general, $t$ depends also on the orientation~\cite{KaxirasPRB16} of this vector relative to the nearest neighbor sites of the atom at $\fvec X_{j,S}$ and at $\fvec X'_{j',S'}$ (Fig.\ref{Fig:Schematic}). Moreover, the general on-site term acquires configuration dependence. We treat this more intricate case in Sec.~\ref{Sec:Extension}.
	Thus, we start with a microscopic tight binding model
	\begin{align}
		H^{SK}_{tb}  = & \sum_{S S'} \sum_{j j'} \sum_{\fvec r_S, \fvec r'_{S'}} t(\fvec X_{j,S} - \fvec X_{j', S'}')   c^{\dagger}_{j, S, \fvec r_S} c_{j', S', \fvec r'_{S'}} , \label{Eqn:MicroH}
	\end{align}
	where the fermion creation and annihilation operators satisfy the anti-commutation relation $\{ c^{\dagger}_{j, S, \fvec r_S},  c_{j', S', \fvec r'_{S'}}  \} = \delta_{j j'} \delta_{S S'} \delta_{\fvec r_S \fvec r'_{S'}}$.
	Because $H^{SK}_{tb}$ is Hermitian,
	\begin{equation}
		\label{Eqn:tHermitian}
		t(\bX) = t^*(-\bX),
	\end{equation}
	and because (spinless) time reversal symmetry is preserved
	\begin{equation}
		\label{Eqn:tTRS}
		t(\bX) = t^*(\bX).
	\end{equation}
One example of a model with $t$ depending only on the separation of the two carbon atoms is the often used Slater-Koster (SK) type model\cite{Ando2001,Uryu2004,SlaterKoster1954,MagaudNL10,KoshinoPRB12,KangVafekPRX} for the carbon $p_z$ orbitals,
\begin{align}
	t(\fvec d) & = V_{pp\pi}^0 e^{- \frac{|\fvec d| - a_0}{\Delta}} \left[  1 - \left( \frac{\fvec d \cdot \hat z}{|\fvec d|} \right)^2 \right] + \nonumber \\
	& V_{pp\sigma}^0 e^{- \frac{|\fvec d| - d_0}{\Delta}}  \left( \frac{\fvec d \cdot \hat z}{|\fvec d|} \right)^2,
	\label{Eqn:SlaterKoster}
\end{align}
where for concreteness $V^0_{pp\pi} = -2.7$eV, $V^0_{pp\sigma} = 0.48$eV, $a_0 =|\fvec \tau_B| = 0.142$nm is the distance between the two nearest-neighbor carbon atoms on the same layer; $d_0 =0.335$nm is the inter-layer distance and the decay length for the hopping is $\Delta = 0.319a_0$.	

In this paper, we will not need to use the specific form of $t$ in (\ref{Eqn:SlaterKoster}). As explained later, we only rely on its fast decay. To obtain the continuum effective Hamiltonian from the microscopic tight binding model (\ref{Eqn:MicroH}), we next write
	\begin{align}
		& H^{SK}_{tb}  =  \ \sum_{S S'} \sum_{j j'} \sum_{\fvec r_S, \fvec r'_{S'}} \int \rmd^2 \fvec r\ \rmd^2 \fvec r'\ \delta(\fvec r - \fvec r_S) \delta(\fvec r' - \fvec r_{S'}')   \nonumber \\
		& \times t(\fvec r + \fvec u_{j,S}(\fvec r) - \fvec r' - \fvec u_{j',S'}(\fvec r')) c^{\dagger}_{j, S, \fvec r} c_{j', S', \fvec r'},
	\end{align}
	interchange the order of summation and integration and apply the ``Dirac comb'' formula
	\[ \sum_{\fvec r_S} \delta(\fvec r - \fvec r_S)= \frac1{A_{mlg}} \sum_{\fvec G} e^{i \fvec G \cdot (\fvec r - \fvec \tau_S)} . \]
	Here $\fvec G=2\pi\left(m_1\fvec a_2-m_2\fvec a_1\right)\times \hat{z}/A_{mlg}$ is the reciprocal lattice vector of the undistorted monolayer graphene, $m_{1,2}$ are integers, and $A_{mlg} = |\fvec a_1 \times \fvec a_2| = \frac{\sqrt{3}}2 a^2$ is the area of the undistorted monolayer graphene unit cell. Since the physically important states come from the vicinity of the Dirac points, we can decompose the fermion fields into two slowly spatially varying fields $\psi$ and $\phi$ multiplied by the fast spatially varying functions from the valley $\fvec K=4\pi\fvec a_1/(3a^2)$ and $\fvec K' = - \fvec K$ as
	\begin{align}
		A_{mlg}^{-1/2} c_{j, S, \fvec r} \simeq e^{i \fvec K \cdot \fvec r} \psi_{j, S}(\fvec r) + e^{- i \fvec K \cdot \fvec r} \phi_{j, S}(\fvec r)  .  \label{Eqn:fastslow}
	\end{align}
	The factor of $A_{mlg}^{-1/2}$ is included to satisfy the anti-commutation relation
	\begin{align}
		& \{\psi_{j,S}(\fvec r), \psi^{\dagger}_{j', S'}(\fvec r') \} = \{\phi_{j,S}(\fvec r), \phi^{\dagger}_{j', S'}(\fvec r') \} \nonumber \\
		= & \delta_{jj'} \delta_{S S'} \delta(\fvec r - \fvec r').
	\end{align}
	The effective Hamiltonian at the valley $\fvec K$ can now be written as
	\begin{align}
		& H_{SK,eff}^\bK  = \frac1{A_{mlg}} \sum_{j j'}\sum_{ S S'}\sum_{\fvec G, \fvec G'} \int \rmd^2 \fvec r\ \rmd^2 \fvec r'\ e^{i \fvec G \cdot (\fvec r - \fvec \tau_S)} e^{-i \fvec G' \cdot (\fvec r' - \fvec \tau_{S'})} \nonumber \\
		&  t(\fvec r + \fvec u_{j,S}(\fvec r) - \fvec r' - \fvec u_{j',S'}(\fvec r'))e^{-i\fvec K\cdot(\fvec r-\fvec r')} \psi^{\dagger}_{j,S}(\fvec r) \psi_{j',S'}(\fvec r'). \label{Eqn:EffH1}
	\end{align}
	The effective Hamiltonian at the valley $\fvec K'$ is related to $H_{SK,eff}^\bK$ by spinless time reversal symmetry.
	
	The hopping amplitude $t$ is a short ranged function that decays exponentially fast as its argument increases beyond a few carbon lattice spacings, while $\psi$ varies slowly over such length scales. In order to take advantage of this fact it is convenient to switch to Eulerian coordinates\cite{Balents19}.
	By doing so, the locality of the effective theory will become manifest\cite{Balents19}. Thus, we now perform a coordinate transformation from $\fvec r$ to $\fvec X^\parallel$, where for each $j$ and $S$ we let the integration variable be $\fvec X^{\parallel}=\fvec r+ \fvec u^\parallel_{j,S}(\fvec r)$, and similarly for the primed variables. We also introduce new $\fvec X^\parallel$-dependent fermion fields as
	\begin{align}
		& \Psi_{j,S}(\fvec X^{\parallel})  = \left| J\left( \frac{\partial \fvec r}{\partial \fvec  X^{\parallel}} \right)  \right|^{\half} \psi_{j, S}(\fvec r)  \nonumber \\
		= & \left| J\left( \frac{\partial (\fvec X^{\parallel} - \fvec U^{\parallel}_{j, S}(\fvec X^{\parallel}))}{\partial \fvec  X^{\parallel}} \right)  \right|^{\half} \psi_{j, S}(\fvec X^{\parallel} - \fvec U^{\parallel}_{j,S}(\fvec X^{\parallel})), \label{Eqn:FieldX}
	\end{align}
	where $J$ is the Jacobi determinant\footnote{Note that this differs from the choice made in Ref.~\cite{Balents19} by a phase factor}.
	As emphasized in Ref.~\cite{Balents19}, $\fvec {U}$ is not small for twisted structures, and the formalism developed here does not make this assumption. However, we do assume that $\fvec {U}$ is smooth i.e. its gradients are small. Therefore, the $\Psi$ fields are also slow. By the property of the Dirac delta functions under the
	change of variables (and because the transformation between $\fvec X^\parallel$ and $\fvec r$ is one-to-one) these fields satisfy the canonical fermion commutation relations
	\begin{equation}\{ \Psi_{j,S}(\fvec X^{\parallel}) , \Psi^{\dagger}_{j',S'}(\fvec X'^{\parallel}) \} = \delta_{jj'} \delta_{S S'} \delta(\fvec X^{\parallel} - \fvec X'^{\parallel}).\end{equation}
	For notational simplicity, we also introduce the symbol $\mathcal{J}_{j,S}(\fvec X^{\parallel}) \equiv   \left| J\left( \frac{\partial \fvec r}{\partial \fvec  X^{\parallel}} \right)  \right|^{\half}$. The effective continuum Hamiltonian now becomes
	\begin{widetext}
		\begin{align}
			& H_{SK,eff}^\bK = \frac1{A_{mlg}} \sum_{j j'} \sum_{S S'} \sum_{\fvec G, \fvec G'} e^{-i (\fvec G \cdot \fvec \tau_s - \fvec G' \cdot \fvec \tau_{S'})} \int \rmd^2 \fvec X^{\parallel}\ \rmd^2 \fvec X'^{\parallel}\ \mathcal{J}_{j,S}(\fvec X^{\parallel}) \mathcal{J}_{j',S'}(\fvec X'^{\parallel}) e^{-i (\fvec G - \fvec K)\cdot \fvec U^{\parallel}_{j,S}(\fvec X^{\parallel})} e^{i (\fvec G' - \fvec K) \cdot \fvec U^{\parallel}_{j', S'}(\fvec X'^{\parallel})} \nonumber \\
			& t(\fvec X^{\parallel} + \fvec U^{\perp}_{j,S}(\fvec X^{\parallel}) - \fvec X'^{\parallel} - \fvec U^{\perp}_{j',S'}(\fvec X'^{\parallel})) e^{i (\fvec G \cdot \fvec X^{\parallel} - \fvec G' \cdot \fvec X'^{\parallel})}
			e^{-i\fvec K\cdot(\fvec X^\parallel-\fvec X'^\parallel)}
			\Psi^{\dagger}_{j,S}(\fvec X^{\parallel}) \Psi_{j',S'}(\fvec X'^{\parallel}). \label{Eqn:EffH2}
		\end{align}
	\end{widetext}
	In order to exploit the short range nature of $t$, we switch to the center-of-mass $\fvec x = \half(\fvec X^{\parallel} + \fvec X'^{\parallel})$ and relative coordinates $\fvec y = \fvec X^{\parallel} - \fvec X'^{\parallel}$. Thus $\int \rmd^2\fvec X^{\parallel}\ \rmd^2 \fvec X'^{\parallel} \ldots= \int \rmd^2\fvec x\ \rmd^2 \fvec y\ldots$, and $e^{i (\fvec G \cdot \fvec X^{\parallel} - \fvec G' \cdot \fvec X'^{\parallel})} = e^{i (\fvec G - \fvec G')\cdot \fvec x} e^{i \half(\fvec G + \fvec G')\cdot \fvec y}$. The integral over $\fvec x$ contains the phase factor  $e^{i (\fvec G - \fvec G')\cdot \fvec x}$ that oscillates strongly over the scale of the monolayer graphene lattice constant $a$ when $\fvec G \neq \fvec G'$, whereas all other factors are smooth functions of $\fvec x$. As a consequence, the integral over $\fvec x$ is negligible as long as $\fvec G \neq \fvec G'$; this collapses the double sum over $\fvec G,\fvec G'$ to a single sum. Moreover, the remaining fields, whether $\Psi$ or $\fvec U$, are smooth and can now be expanded in powers of gradients e.g. $\Psi(\fvec x-\frac{1}{2}\fvec y)\simeq \Psi(\fvec x)-\frac{1}{2}\fvec y\cdot\nabla_{\fvec x} \Psi(\fvec x)+\frac{1}{8}\left(\fvec y\cdot\nabla_{\fvec x}\right)^2\Psi(\fvec x)\ldots$, because powers of $\fvec y$ are compensated by the exponential decay of $t$ at large $\fvec y$, effectively confining $\fvec y$ to small values.
	Changing $\fvec G$ to $- \fvec G$ and to the first order in gradients, we obtain the main result of this section
	\begin{widetext}
		\begin{eqnarray}
			H^{\bK}_{SK,eff}
			&&\simeq\frac{1}{A_{mlg}} \sum_{S,S'} \sum_{jj'} \sum_{\fvec G} e^{i\fvec G \cdot(\fvec{\tau}_S-\fvec{\tau}_{S'})} \int \rmd^2 \fvec x	\mathcal{J}_{j,S}(\fvec x)
			\mathcal{J}_{j',S'}(\fvec x)	e^{i(\fvec G + \fvec K)\cdot \left(\fvec U^\parallel_{j,S}(\fvec x) - \fvec U^\parallel_{j',S'}(\fvec x) \right)}
			\nonumber\\
			& &\int \rmd^2\fvec y e^{-i(\fvec G + \fvec K) \cdot \fvec y} e^{i\frac{\fvec y}{2} \cdot \nabla_{\fvec x} \left( \fvec U^\parallel_{j,S}(\fvec x) + \fvec U^\parallel_{j',S'}(\fvec x) \right) \cdot (\fvec G + \fvec K)}	t\left[ \fvec y + \fvec U^\perp_{j,S}\left(\fvec x\right) - \fvec U^\perp_{j',S'}(\fvec x)\right]	\nonumber\\
			& & \times	\left[ \Psi^\dagger_{j,S}(\fvec x) \Psi_{j',S'}(\fvec x) + \frac{\fvec y}2 \cdot\left( \left(\nabla_{\fvec x} \Psi^\dagger_{j,S}(\fvec x) \right) \Psi_{j',S'}(\fvec x) - \Psi^\dagger_{j,S}(\fvec x) \nabla_{\fvec x} \Psi_{j',S'}(\fvec x)\right)
			\right].   \label{Eqn:EffContH}
		\end{eqnarray}
	\end{widetext}
	Extension to higher order gradients is straightforward (see Appendix \ref{App:quadratic}).  We analyze the accuracy of this formula for the Slater-Koster like models of twisted bilayer graphene~\cite{MagaudNL10, KoshinoPRB12}, including in-plane lattice relaxation, by comparing the low energy continuum and tight-binding spectra in the companion paper, by also including a second order gradient terms in the intra-layer part of the effective Hamiltonian. Analogous formula is derived in the next two sections for model in Ref.~\cite{KaxirasPRB16} that includes dependence of $t$ on the relative orientation of the intra-layer nearest neighbor sites and the vector connecting two inter-layer sites $\fvec X_{j,S} - \fvec X'_{j', S'}$.

	The inter-valley scattering terms are negligibly small. This can be seen by a direct substitution of (\ref{Eqn:fastslow}), following the analysis above, and noticing that the vector $2\fvec K$ is {\it not} a reciprocal lattice vector $\bG$, while $3\fvec K$ is. For the intervalley scattering we therefore need to compensate for the missing $\bK$ using terms of order $\sim\bG \cdot \partial_\mu \fvec{U}$. Because for rigid twist angle $\theta$, $\partial_\mu \fvec{U}\sim \theta$ and because the relaxed atomic configuration is smooth on the moire scale $L_m$, more generally $\partial_\mu \fvec{U}\sim a/L_m\ll 1$. This forces us to either go to very high $\bG\sim \fvec K L_m/a$ for which the Fourier transform of $t$ is exponentially small, or, for smaller $\bG$ to extract the Fourier component of terms of the form $e^{i(\bG+\bK)\cdot \fvec{U}(\fvec x)}$ at $\fvec K$. Upon Fourier expanding $\fvec{U}(\fvec x)$, the function $e^{i(\bG+\bK)\cdot \fvec{U}(\fvec x)}$ can be thought of as a product of generating functions for the Bessel functions. While non-zero, Fourier component of $e^{i(\bG+\bK)\cdot \fvec{U}(\fvec x)}$ at $\fvec K$ corresponds to Bessel functions at high indices with arguments set by the $\partial_\mu\fvec{U}(\fvec x)$, which are exponentially small. Inspecting the tight binding spectra analyzed in the companion paper, which contain the intervalley scattering terms, and comparing them with the spectra obtained from the continuum models which neglect them, indeed justifies neglecting the intervalley scattering terms over the experimentally relevant energy scale.
	
	Different from the earlier works~\cite{KoshinoPRB20}, our derivation does not distinguish the intralayer and interlayer Hamiltonians of the continuum theory. Consequently, the inter-layer tunneling obtained from Eqn.~\ref{Eqn:EffContH} depends not only on the asymmetric lattice displacement $\fvec U^{\parallel}_{t, S} - \fvec U^{\parallel}_{b, S'}$, but also on the gradient of the symmetric part $\fvec \nabla (\fvec U^{\parallel}_{t, S} + \fvec U^{\parallel}_{b, S'})$.  
	
\subsection{Bond Orientation Dependent Hopping}\label{Sec:Extension}
In the previous section we derived the effective continuum Hamiltonian when the hopping depends only on the separation of two carbon atoms $\fvec X_{j, S} - \fvec X'_{j', S'}$, as is the case for Slater-Koster type models~\cite{Ando2001,Uryu2004,SlaterKoster1954,MagaudNL10,KoshinoPRB12,KangVafekPRX}. In such models, the Wannier states are essentially the atomic $p_z$ orbitals on each carbon atom, and therefore the full azimuthal symmetry is retained making the inter-layer hoppings in-plane isotropic (see Eq.\ref{Eqn:SlaterKoster}), with no dependence on the three nearest neighbor bond vectors $\fvec n^{(\alpha)}_{j,S}(\fvec X)$ at the position of $\fvec X_{j,S}$ where $\alpha = 1,2$, or $3$ (see Fig.~\ref{Fig:Schematic}); and similarly no dependence on $\fvec n^{(\alpha')}_{j',S'}(\fvec X')$. In the more detailed microscopic model derived from DFT determined Wannier states of the monolayer (and untwisted bilayer) graphene's conduction and valance bands~\cite{KaxirasPRB16}, the localized state indeed has a dominant $p_z$ character, but the azimuthal symmetry is lost due to the trigonal crystal field of the neighboring atoms. The localized state is therefore a superposition of several lattice harmonics with angular momenta $L_z = 0$, $3$, $6$, etc.  As a consequence, the inter-layer hopping part of $t$ acquires the dependence on the relative orientation of the atomic separation vector $\fvec X_{j, S} - \fvec X'_{j', S'}$ and $\fvec n^{(\alpha)}_{j,S}(\fvec X)$, $\fvec n^{(\alpha')}_{j',S'}(\fvec X')$.

Here we generalize Eq.(\ref{Eqn:EffContH}) to include such effects on the effective continuum Hamiltonian.
In Eq.(\ref{Eqn:MicroH}) therefore replace
\begin{eqnarray}
    && t(\fvec X_{j,S} - \fvec X_{j', S'}')\rightarrow\nonumber\\
    && t(\fvec X_{j,S} - \fvec X_{j', S'}',  \{\fvec n^{(\alpha)}_{j, S}(\fvec X) \},  \{\fvec n_{j', S'}^{(\alpha)}( \fvec X')\} ),
\label{Eqn:tBondOrient}\end{eqnarray}
where the $\{\}$ denotes the dependence on each term in the set, i.e. $\alpha=1,2,3$.
Next, from the definition of the nearest neighbor vectors we can write
\begin{eqnarray}\label{Eqn:nVec} \fvec n^{(\alpha)}_{j, S}(\fvec X) = \fvec X^{\parallel}_{j, \bar{S}}(\fvec r_S + \fvec \delta_{S}^{(\alpha)}) - \fvec X^{\parallel}_{j,S}(\fvec r_S),
\end{eqnarray}
where $\fvec \delta_S^{(\alpha)}$ are the three nearest neighbor bond vectors of the undistorted lattice, that can be expressed as 
\begin{eqnarray}
\fvec \delta_S^{(\alpha)} &=& R(2\pi(\alpha-1)/3) \fvec \delta_S^{(1)},\\
\fvec \delta_S^{(1)} &=&  \fvec \tau_{\bar{S}} - \fvec \tau_S\equiv \fvec\delta_S,
\end{eqnarray} 
where $R(\omega)$ is the two-dimensional rotation matrix with the angle $\omega$
\begin{eqnarray}
    R(\omega)=\left(\begin{array}{cc} \cos\omega & -\sin\omega \\
    \sin\omega & \cos\omega\end{array}\right).
\end{eqnarray}
With our choice of the coordinate system, $\fvec \delta_A^{(1)}=\fvec\tau_B=\frac{1}{3}(\fvec a_1+\fvec a_2)$, $\fvec \delta_A^{(2)}=\frac{1}{3}\left(\fvec a_2 -2\fvec a_1\right)$, $\fvec \delta_A^{(3)}=\frac{1}{3}\left(\fvec a_1 -2\fvec a_2\right)$, and $\fvec\delta_B^{(\alpha)}=-\fvec\delta_A^{(\alpha)}$.

We will consider only the atomic configurations which are varying smoothly not only within a sublattice but also between the two sublattices.
All configurations examined in the companion paper are of this type.
Therefore, we can drop the $S$ subscript in Eq.(\ref{Eqn:EulerCoord})
\begin{equation} \fvec X_{j,S}(\fvec r) \equiv \fvec X_j(\fvec r) = \fvec r + \fvec U^{\parallel}_j(\fvec X_j) + \fvec U^{\perp}_j(\fvec X_j).\end{equation}
Correspondingly, the bond vectors in Eq.(\ref{Eqn:nVec}) become $\fvec n^{(\alpha)}_{j,S}(\fvec X) = \fvec X_j^{\parallel}(\fvec r_S + \fvec \delta_S^{\alpha}) - \fvec X_j^{\parallel}(\fvec r_S)$. Introducing a continuum variable $\fvec r$  and changing the integration variable to $\fvec X^\parallel=\fvec r+\fvec u^\parallel_j(\fvec r)$ for each $j$ and $S$, as in the previous section makes the bond vectors $\fvec n^{(\alpha)}_{j,S}$ a function of $\fvec{X}^\parallel$.
Therefore, $t$ in Eq.(\ref{Eqn:EffH2}) gains an additional dependence on $\fvec n_{j,S}^{(\alpha)}(\fvec X^{\parallel})$ and $\fvec n_{j',S'}^{(\alpha)}(\fvec X'^{\parallel})$.
For a smooth atomic displacement fields we can then write
\begin{eqnarray}
    \fvec n_{j,S}^{(\alpha)}(\fvec X^{\parallel}) &=&
    \fvec \delta_S^{\alpha} +
    \fvec u_j^\parallel\left(\fvec X^\parallel-\fvec U_j^\parallel\left(\fvec X^\parallel\right)+\fvec \delta_S^{\alpha} \right)-\fvec U_j\left(\fvec{X}^\parallel\right)\nonumber\\
    &&\simeq \fvec \delta_S^{\alpha}+ \delta_{S,\mu}^{\alpha}\frac{\partial \fvec u_j^\parallel}{\partial r_\mu}\simeq \fvec \delta_S^{\alpha}+ \delta_{S,\mu}^{\alpha}\frac{\partial \fvec U^\parallel_{j}\left(\fvec X^\parallel\right)}{\partial X^\parallel_\mu},  \label{Eqn:NNBond}
\end{eqnarray}
because
\begin{eqnarray}
  &&  \frac{\partial u_{j,\nu}^\parallel}{\partial r_\mu}=
        \frac{\partial \left(X_{\nu}^\parallel-r_\nu\right)}{\partial r_\mu}=\frac{\partial X_{\nu}^\parallel}{\partial r_\mu}-\delta_{\mu\nu}=\left(\frac{\partial \fvec r}{\partial \fvec X}\right)_{\nu\mu}^{-1}-\delta_{\mu\nu}\nonumber\\
        &&=\left(\frac{\partial \left(\fvec X^\parallel-\fvec U^\parallel_j\right)}{\partial \fvec X^\parallel}\right)_{\nu\mu}^{-1}-\delta_{\mu\nu}\simeq \frac{\partial U^\parallel_{j,\nu}}{\partial X^\parallel_\mu}.
\end{eqnarray}
By going to center-of-mass and relative coordinates,  and keeping only term up the first order derivative of $\fvec U^{\parallel}$, we find
\begin{align}
	& \fvec n^{(\alpha)}_{j, S}\left(\fvec x \pm \half \fvec y\right) \simeq \fvec \delta_S^{(\alpha)}+ \delta_{S,\mu}^{(\alpha)}\frac{\partial \fvec U^\parallel_{j}\left(\fvec x\right)}{\partial x_\mu}.
\end{align}
Following the arguments that led to the Eq.(\ref{Eqn:EffH2}), we find that $H^\bK_{eff}$ can be obtained from Eq.(\ref{Eqn:EffContH}) if for each layer index $j$, $j'$, we drop the sublattice index on $\fvec U$ i.e. we replace $\fvec U^{\parallel,\perp}_{j,S}(\fvec x)\rightarrow  \fvec U^{\parallel,\perp}_{j}(\fvec x)$ and similarly for $j'$, $S'$, and we replace
\begin{eqnarray}\label{Eqn:tReplacement}
&&    t\left[\fvec d_{S,S'}\right]\rightarrow \\
  &&  t\left[\fvec d,
	\left\{\fvec \delta_S^{(\alpha)}+ \delta_{S,\mu}^{(\alpha)}\frac{\partial \fvec U^\parallel_{j}\left(\fvec x\right)}{\partial x_\mu} \right\},
\left\{\fvec \delta_{S'}^{(\alpha)}+ \delta_{S',\mu}^{(\alpha)}\frac{\partial \fvec U^\parallel_{j'}\left(\fvec x\right)}{\partial x_\mu}\right \} \right].\nonumber
\end{eqnarray}
With these replacements, Eq.(\ref{Eqn:EffContH}) gives the effective continuum Hamiltonian for the bond orientation dependent inter-layer hopping for an arbitrary, sublattice independent, smooth atomic deformation. The additional configuration dependent on-site term is discussed in the next subsection.

\subsection{Bond Dependent On-Site Energy}
The onsite terms in the tight binding model need to be considered separately because in practice they may not be accounted for accurately by the continuous interpolation function $t$ in the expression (\ref{Eqn:tReplacement}). We assume that the difference between the full configuration dependence of the on-site term and the contribution from $t$ at $\fvec d=0$ can be approximated by the form
\begin{align}
		H_{onsite} = & \sum_{j, S} \sum_{\fvec r_S} \epsilon\left( \left\{ |\fvec  n_{j, S}^{(\alpha)}(\fvec X_{j, S})| \right\} \right) c^{\dagger}_{j, S, \fvec r_S} c_{j, S, \fvec r_S} , \label{Eqn:HOnSite}
\end{align}
where the onsite energy $\epsilon$ is assumed to depend on the length of three nearest bonds $\fvec n_{j, S}^{(\alpha)}(\fvec X)$, defined in Eq.~\ref{Eqn:NNBond}. Applying the same methods, we can write
\begin{align}
    & H_{onsite}= \nonumber \\
      & \frac1{A_{mlg}} \sum_{j,S, \fvec G} \int \rmd^2 \fvec r\ e^{i \fvec G \cdot (\fvec r - \fvec \tau_S)} \epsilon\left( \left\{ \fvec  n_{j, S}^{(\alpha)}(\fvec X_{j, S}) \right\} \right) c^{\dagger}_{j, S, \fvec r} c_{j, S, \fvec r}.
\end{align}
Next, we introduce the field operator $\psi_{j, S}(\fvec r)$ via the Eq.~\ref{Eqn:fastslow} in order to obtain the correction to the effective Hamiltonian at the valley $\fvec K$ from the on-site term. Changing the integration variable $\fvec r$ to $\fvec X^{\parallel}$ introduces the Jacobi determinant $|J(\partial \fvec r/ \partial \fvec X^{\parallel})|$. As shown in Eq.~\ref{Eqn:FieldX}, this factor is absorbed by the redefinition of the field operator $\Psi_{j, S}(\fvec X^{\parallel})$. In addition, $e^{i \fvec G \cdot \fvec r} = e^{i \fvec G \cdot (\fvec X^{\parallel} - U^{\parallel}(\fvec X^{\parallel}))}$. Thus, the onsite term at the valley $\fvec K$ can be written as
\begin{align}
    H^{\bK}_{onsite} =  & \sum_{j, S, \fvec G} e^{-i \fvec G \cdot \fvec \tau_S} \int \rmd^2 \fvec X^{\parallel}\ e^{i \fvec G \cdot (\fvec X^{\parallel} - U^{\parallel}(\fvec X^{\parallel}))} \nonumber \\
    & \epsilon\left( \left\{ |\fvec  n_{j, S}^{(\alpha)}(\fvec X^{\parallel}) |\right\} \right) \Psi^{\dagger}_{j,S}(\fvec X^{\parallel}) \Psi_{j,S}(\fvec X^{\parallel}). \label{Eqn:OnSiteGSum}
\end{align}
If $\fvec G \neq 0$, the factor $e^{i \fvec G \cdot \fvec X^{\parallel}}$ oscillates around zero on the scale of the carbon-carbon distance and because it multiplies much more slowly varying functions of $\fvec X^{\parallel}$ the integral vanishes. Therefore, we can keep only the term with $\fvec G = 0$ in the above sum and obtain
\begin{align}
    H^{\bK}_{eff,onsite} =  & \sum_{j, S}  \int \rmd^2 \fvec x\
    \epsilon\left( \left\{ |\fvec  n_{j, S}^{(\alpha)}(\fvec x) |\right\} \right) \Psi^{\dagger}_{j,S}(\fvec x) \Psi_{j,S}(\fvec x).
\end{align}
To the linear order of gradients of $\fvec U^{\parallel}$, the length of the distorted nearest neighbor bond is
\begin{align}
    |\fvec  n_{j, S}^{(\alpha)}(\fvec x) | \approx |\fvec \delta_S^{\alpha}| + \delta_{S, \mu}^{\alpha} \frac{\partial U^{\parallel}_{j,\mu}}{\partial x_{\nu}} \delta_{S, \nu}^{\alpha}/|\fvec \delta_S^{\alpha}|,
\end{align}
where the length of the undistorted nearest neighbor bond vectors is the same $|\fvec \delta_S^{\alpha}| = a/\sqrt{3}$. Thus, to leading order gradient expansion, the onsite energy $\epsilon$ is
\begin{equation}
\epsilon\left( \left\{ |\fvec  n_{j, S}^{(\alpha)}(\fvec x) |\right\} \right) 
    \approx  \epsilon\left( \frac{a}{\sqrt{3}} \right) + \frac{\sqrt{3}}{a}\sum_{\alpha = 1}^3 \frac{\partial \epsilon_0}{\partial |\delta_S^{\alpha}|} \delta_{S, \mu}^{\alpha} \frac{\partial U^{\parallel}_{j,\mu}}{\partial x_{\nu}} \delta_{S, \nu}^{\alpha}.
\end{equation}
Due to $C_3$ and space inversion symmetries, $\partial \epsilon/\partial |\delta_S^{\alpha}|$ is independent of $\alpha$ and $S$. In addition, $\sum_{\alpha} \delta_{S, \mu}^{\alpha} \delta_{S, \nu}^{\alpha} = \delta_{\mu\nu}a^2/2$. Introducing $\epsilon_0 = \epsilon(a/\sqrt{3})$ and $\kappa = \frac{\sqrt{3}}2 a \left( \partial \epsilon/\partial |\delta_S^{\alpha}| \right)$, we obtain 
\begin{align}
    \epsilon\left( \left\{ |\fvec  n_{j, S}^{(\alpha)}(\fvec x) |\right\} \right) \approx \epsilon_0 + \kappa \fvec \nabla \cdot \fvec U_j^{\parallel} \ .
\end{align}
Therefore, the contribution of the on-site term to the effective continuum Hamiltonian at $\fvec K$ is
\begin{align}
    H_{eff,onsite}^{\bK} = \sum_{j, S} \int \rmd^2\fvec x\ \left( \epsilon_0 + \kappa \fvec \nabla \cdot \fvec U^{\parallel}_j \right) \Psi_{j, S}^{\dagger}(\fvec x) \Psi_{j, S}(\fvec x), \label{Eqn:HEffOnSite}
\end{align}
thus correcting the value of the deformation potential obtained from $t$ alone. 

\subsection{Bond Orientation Dependent Microscopic Model of Ref.~\cite{KaxirasPRB16}}

In the derivation above, we allow for a general form of the hopping, depending on all the nearest neighbor bond vectors $\fvec n^{(\alpha)}_{j,S}$ and $\fvec n^{(\alpha)}_{j',S'}$. The model of Ref.~\cite{KaxirasPRB16} was derived for configurations which are locally $C_3$ symmetric, i.e.~all three bond vectors $\fvec n^{(\alpha)}_{j,S}$ are equivalent to each other, as are the three bond vectors $\fvec n^{(\alpha)}_{j', S'}$. In this case, the bond dependence can be simplified because the hopping is the same for each one of the three bond vectors. With an eye towards generalizing to smooth lattice distortions which lead to a (small) violation of the local $C_3$ symmetry, we write the formula for the hoppings in Eq.(\ref{Eqn:tBondOrient}) as
\begin{align}
	& t(\fvec X_{j,S} - \fvec X_{j', S'}',  \{\fvec n^{(\alpha)}_{j, S}(\fvec X) \},  \{\fvec n_{j', S'}^{(\alpha)}( \fvec X')\} )= \nonumber \\
  & \frac19 \sum_{\alpha = 1}^3 \sum_{\alpha' = 1}^3 t^{jj'}_{sym}(\fvec X_{j,S} - \fvec X_{j', S'}',  \fvec n^{(\alpha)}_{j, S}(\fvec X),  \fvec n_{j', S'}^{(\alpha')}( \fvec X')\} ), \label{EqnS:InterHopping}
\end{align}
where $t^{jj'}_{sym}$ is the hopping function of Ref.~\cite{KaxirasPRB16} when the configuration is locally $C_3$ symmetric.
For the intra-layer hopping
\begin{equation}
	 t_{sym}^{j = j'}(\fvec X_{j,S} - \fvec X_{j', S'}',  \fvec n^{(\alpha)}_{j, S}(\fvec X),  \fvec n_{j', S'}^{(\alpha')}( \fvec X')\} =\tilde{V}_0(y), \label{Eqn:KaxirasIntra}  
	 \end{equation}
	 where
	 \begin{equation}
	 \tilde{V}_0(y) =    \tilde{\lambda}_0 e^{-\tilde{\xi}_0 \left(y/a\right)^2} \cos\left(\tilde{\kappa}_0 \frac{y}{a}\right) +  \tilde{\lambda}_1 \frac{y^2}{a^2} e^{- \tilde{\xi}_1 (y/a - \tilde{x}_1)^2},
	\end{equation}
where $\fvec y=\fvec X^\parallel_{j,S} - \fvec X'^{\parallel}_{j',S'}$ is the in-plane projected separation vector, $y=|\fvec y|$ is its magnitude. The intra-layer hopping with $j = j'$ is rotationally isotropic, depending only on $y$. Note that its explicit formula is not provided by Ref.~\cite{KaxirasPRB16}, in which the hopping constants are listed only for discrete values of $y$, i.e. for distances of several pairs of carbon atoms on the undistorted monolayer graphene lattice. To obtain the values of the hopping constants with arbitrary $y$, we fit these hopping constants with the formula in Eq.~\ref{Eqn:KaxirasIntra} and extract the parameters that are listed in the left table of Table~\ref{Tab:KaxirasHopping}.

\begin{table}[t]
	\centering
	\begin{tabular}{|c|c|c|} \hline
					   $i$ &  $0$	& $1$ \\ \hline
	$\tilde{\lambda}_i/\mathrm{eV}$&  $-18.4295$ 	& $-3.7183$   \\ \hline
	$\tilde{\xi}_i$    &  $1.2771$ 		&  $6.2194$  \\ \hline 
	$\tilde{x}_i$      &           		&  $0.9071$   \\ \hline
	$\tilde{\kappa}_i$ &   $2.3934$ 	&    \\ \hline
\end{tabular}
\
	\begin{tabular}{|c|c|c|c|c|c|} \hline
		           $i$ &   $0$      	& $3$ 	& $6$    \\		\hline
	    $\lambda_i/\mathrm{eV}$&  $0.3155$    	& $-0.0688$ &  $-0.0083$  \\ \hline
	    $\xi_i$    &  $1.7543$    	& $3.4692$  &  $2.8764$  \\ \hline 
		$x_i$      &    			& $0.5212$ 	& $1.5206$ 	 \\		\hline
		$\kappa_i$ &   $2.0010$     &   		&  $1.5731$   \\ \hline
	\end{tabular}
	\caption{Parameters of the formula for the intra-layer hopping with (left) $j = j'$ in Eq.~\ref{Eqn:KaxirasIntra} and inter-layer hopping (right, and from Ref.\cite{KaxirasPRB16}) $j \neq j'$ in Eq.~\ref{Eqn:KaxirasInter}. }
	\label{Tab:KaxirasHopping}
\end{table}

In the locally $C_3$ symmetric case, the inter-layer part of the $t^{jj'}_{sym}$ depends only on two bond vectors, one at $\fvec X_{j,S}$ and another one at $\fvec X'_{j',S'}$, as
\begin{eqnarray}
 t^{j\neq j'}_{sym}(&&\fvec X_{j,S} - \fvec X'_{j',S'}, \fvec n_{j,S}, \fvec n_{j',S'}) = V_0(y) + \nonumber \\
 &&V_3(y) \left( \cos(3\theta_{12}) + \cos(3\theta_{21})  \right)  +  \nonumber\\
 &&V_6(y) \left( \cos(6\theta_{12}) + \cos(6\theta_{21})  \right). \label{Eqn:KaxirasInter}
\end{eqnarray}
The explicit formulas for  $V_i(y)$ are presented in Ref.~\cite{KaxirasPRB16} and we include them here for completeness
\begin{eqnarray}
	 V_0(y) &=&  \lambda_0 e^{-\xi_0 (y/a)^2} \cos\left(\kappa_0 \frac{y}{a}\right),  \\
	 V_3(y)  &=&  \lambda_3 \frac{y^2}{a^2} e^{-\xi_3 (y/a - x_3)^2}, \\
	V_6(y)  &=&  \lambda_6 e^{-\xi_6 (y/a - x_6)^2} \sin\left(\kappa_6 \frac{y}{a}\right).
\end{eqnarray}
The parameters are specified in Table \ref{Tab:KaxirasHopping}.

The variables $\theta_{12}$ and $\theta_{21}$ in Eq.(\ref{Eqn:KaxirasInter}) are the angles between $\fvec y$ and the nearest neighbor bond vectors on two layers, i.e.
\begin{align}
	&\theta_{12} 
	 = \cos^{-1}\left( -\frac{\fvec y \cdot \fvec n_{j,S}}{y |\fvec n_{j,S}|} \right) = \theta_{\fvec y} - \theta_{j,S} + \pi, \\
	&\theta_{21}  = \cos^{-1} \left( \frac{\fvec y \cdot \fvec n_{j',S'}}{y |\fvec n_{j',S'}|}  \right) = \theta_{\fvec y} - \theta_{j',S'} \ .
\end{align}
In the above we defined $\theta_{\fvec y}$ to be the angle between the separation vector $\fvec y$ and the $x$ axis, and $\theta_{j,S}$ ($\theta_{j',S'}$) to be the angle between the bond vector $\fvec n_{j, S}$ ($\fvec n_{j', S'}$) and the $x$ axis. $\theta_{j,S}^{(\alpha)}$ ($\theta_{j',S'}^{(\alpha)}$) is introduced similarly but with the superscript $\alpha$ to distinguish the angle of different bond vectors. In the absence of the lattice distortion (e.g. as for a rigid twist), the three in-plane nearest neighbors of a carbon atom are $C_3$ symmetric about the carbon atom, and $\theta^{(\alpha)}_{j, S} = \theta_{j, S}^{(1)} + 2\pi(\alpha - 1)/3$. Therefore, the angles $\theta_{12}$ and $\theta_{21}$ could differ by $2\pi/3$ if choosing a different nearest neighbor bond, leading to the same $\cos(3 m \theta_{12})$ and $\cos(3 m \theta_{21})$ with $m$ being an integer.
Therefore, without distortions each term in the sum on the right hand side of (\ref{EqnS:InterHopping}) is the same and the sum is redundant.
In the presence of the lattice relaxation, however, the local $C_3$ symmetry is in general broken and the bond vectors become inequivalent. In order to generalize $t$ to include such slowly varying atomic displacements, we use the formula (\ref{EqnS:InterHopping}).
With the local $C_3$ symmetry broken, the difference between the angles $\theta^{(\alpha)}_{j, S}$ deviates from $\pm 2\pi/3$. For smooth lattice deformation the deviation is small. To obtain this deviation, we write $\fvec n_{j, S}^{(\alpha)} = \fvec \delta_S^{(\alpha)} + \delta  \fvec n_{j, S}^{(\alpha)}$ with $\delta \fvec n_{j, S}^{(\alpha)} = \delta^{(\alpha)}_{S,\mu}\frac{\partial \fvec U_j^{\parallel}}{\partial x_{\mu}} $  and expand the angle $\theta_{j,S}^{(\alpha)}$ to the linear order of the derivatives of $\fvec U^{\parallel}_j$ as
\begin{align}
	& \theta_{j,S}^{(\alpha)} = \theta_{\fvec \delta_S^{(\alpha)}} + \delta \theta^{(\alpha)}_{j,S},  \\
	  & \delta \theta_{j,S}^{(\alpha)} = \frac{(\hat z \times \fvec \delta_S^{(\alpha)})\cdot \delta \fvec n_{j,S}^{(\alpha)}}{|\fvec \delta_S^{(\alpha)}|^2} =  \frac{\epsilon_{\mu\nu}}{|\fvec \delta_S^{(\alpha)}|^2}  \delta_{S,\mu}^{(\alpha)} \frac{\partial U_{j,\nu}^{\parallel}}{\partial x_{\rho}} \delta_{S,\rho}^{(\alpha)}.
\end{align}
$\theta_{\fvec \delta_B^{(\alpha)}}=\theta_{\fvec \delta_A^{(\alpha)}}+\pi$, and for our choice of the coordinate system, $\theta_{\fvec \delta_A^{(1)}}=\pi/6$, $\theta_{\fvec {\delta}_A^{(2)}}=\pi/6+2\pi/3$, $\theta_{\fvec \delta_A^{(3)}}=\pi/6-2\pi/3$.

For the inter-layer part we therefore introduce derivatives of $t^{j j'}_{sym}$ with respect to the angles as
\begin{align}
	& t^{(1)}_{j\neq j',S}(\fvec y) = \left. \frac{\partial t^{j\neq j'}_{sym}}{\partial \theta_{j, S}} \right|_{\theta_{j,S} = \theta_{\fvec \delta_S}} = \\
	 &  
		-3 V_3(y) \sin(3(\theta_{\fvec y} - \theta_{\fvec \delta_S})) + 6 V_6(y) \sin(6(\theta_{\fvec y} - \theta_{\fvec \delta_S}))   , \nonumber \\
	& t^{(2)}_{j\neq j',S'}(\fvec y) = \left. \frac{\partial t^{j\neq j'}_{sym}}{\partial \theta_{j',S'}} \right|_{\theta_{j',S'} = \theta_{\fvec \delta_{S'}}} = \nonumber \\
	& 	3 V_3(y) \sin(3(\theta_{\fvec y} - \theta_{\fvec \delta_{S'}})) + 6 V_6(y) \sin(6(\theta_{\fvec y} - \theta_{\fvec \delta_{S'}}))\end{align}
and vanishing for the intra-layer part
\begin{equation}
	t^{(1)}_{j=j',S}(\fvec y)=t^{(2)}_{j=j',S'}(\fvec y)=0.
\end{equation}
The above expressions are clearly independent under $\theta_{\fvec \delta_S}\rightarrow \theta_{\fvec \delta_S}\pm2\pi/3$, and therefore it does not matter which $\theta_{\fvec \delta^{(\alpha)}_S}$ is substituted for $\theta_{\fvec \delta_S}$.
Thus, combining with Eq.~\ref{Eqn:HEffOnSite}, for an arbitrary smooth lattice deformation, the effective continuum Hamiltonian for the lattice model of Ref.~\cite{KaxirasPRB16} is 
\begin{widetext}
	\begin{eqnarray}
		&&	H^{\bK}_{eff} \simeq \frac{1}{A_{mlg}} \sum_{S,S'} \sum_{jj'} \sum_{\fvec G} e^{i\fvec G \cdot(\fvec{\tau}_S-\fvec{\tau}_{S'})} \int \rmd^2 \fvec x\	\mathcal{J}_{j}(\fvec x)
		\mathcal{J}_{j'}(\fvec x)	e^{i(\fvec G + \fvec K)\cdot \left(\fvec U^\parallel_{j}(\fvec x) - \fvec U^\parallel_{j'}(\fvec x) \right)}
		\int \rmd^2\fvec y e^{-i(\fvec G + \fvec K) \cdot \fvec y}
		\nonumber\\
		&& \times e^{i\frac{\fvec y}{2} \cdot \nabla_{\fvec x} \left( \fvec U^\parallel_{j}(\fvec x) + \fvec U^\parallel_{j'}(\fvec x) \right) \cdot (\fvec G + \fvec K)}	
		\left(t^{j j'}_{sym}\left[ \fvec y + \fvec U^\perp_{j}\left(\fvec x\right) - \fvec U^\perp_{j'}(\fvec x),
		\fvec \delta_S, \fvec \delta_{S'} \right]	
		+ t^{(1)}_{j j', S}(\fvec y) \frac13 \sum_{\alpha = 1}^3  \delta \theta_{j,S}^{(\alpha)} + t^{(2)}_{j j', S'}(\fvec y) \frac13\sum_{\alpha' = 1}^3 \delta \theta_{j', S'}^{(\alpha')}
		\right)\nonumber\\
		&&  \times	\left[ \Psi^\dagger_{j,S}(\fvec x) \Psi_{j',S'}(\fvec x) + \frac{\fvec y}2 \cdot\left( \left(\nabla_{\fvec x} \Psi^\dagger_{j,S}(\fvec x) \right) \Psi_{j',S'}(\fvec x) - \Psi^\dagger_{j,S}(\fvec x) \nabla_{\fvec x} \Psi_{j',S'}(\fvec x)\right)
		\right] \nonumber \\
		&& + \sum_{j, S} \int \rmd^2\fvec x\ \left( \epsilon_0 + \kappa \fvec \nabla \cdot \fvec U^{\parallel}_j(\fvec x) \right) \Psi_{j, S}^{\dagger}(\fvec x) \Psi_{j, S}(\fvec x) \  . 
		\label{Eqn:EffContHCorrection}
	\end{eqnarray}
\end{widetext}
The comparison between the continuum and tight-binding spectra for the model of Ref.~\cite{KaxirasPRB16} for rigid twist as well as for the (relaxed) atomic configurations obtained from solving continuum elastic theory for twisted bilayer are shown in the companion paper.

	\section{Discussion}
	\label{sec:discussion}
In our derivation of the continuum effective Hamiltonians $H_{eff}^\bK$ for graphene bilayers, we have not made use of symmetries. Although this might seem reasonable given that we are considering arbitrary smooth inhomogeneous atomic configurations which would remove any remaining symmetries, as pointed out by Balents~\cite{Balents19}, the form of the leading order terms in the effective Hamiltonian can nevertheless be further constrained. That is because $H_{eff}^\bK$ must be invariant under symmetry operations of the undistorted lattice (i.e. AA-stacked bilayer) that leave a valley invariant if we {\it simultaneously} transform the fermion operators {\it and} the atomic displacement fields\cite{Balents19}.

Although we postpone the detailed analysis of the symmetry, here and in the Appendix \ref{App:symmetry} we would like to highlight some of its consequences. The symmetries of interest to us will be  $C_3$, $C_2\mathcal{T}$, $C_{2x}$, and $\mathcal{R}_y$. Here $C_3$ is the three-fold rotation along $z$ axis. $C_2\mathcal{T}$ is the time reversal followed by the two-fold rotation along $z$ axis with the two sublattices interchanged. $\mathcal{R}_y$ is the mirror reflection along $xz$ plane bisecting the nearest neighbor carbon bond so that $(x, y, z) \rightarrow (x, - y, z)$ and the sublattice index $A \leftrightarrow B$. And $C_{2x}$ is $\mathcal{R}_y$ followed by the interchange of the two layers i.e. followed by the  $xy$ plane mirror reflection half-way between the layers $\mathcal{R}_z$. 

The consequences of $C_2\mathcal T$ and $\mathcal{R}_y$ at $\bG=0$ for the contact interlayer tunneling term --independent of the spatial gradients of the atomic displacement i.e. to zeroth order in $\nabla_{\fvec x}\fvec U$-- were worked out in Ref.\cite{Balents19}.  There it was shown that, when combined with $C_3$, only two independent real parameters are allowed for the first shell of wavectors $\bG=0,-4\pi \fvec a_2\times \hat{z}/\sqrt{3}a^2, 4\pi(\fvec a_1-\fvec a_2)\times \hat{z}/\sqrt{3}a^2$. Physically, these correspond to the interlayer tunneling through the AA region and the AB region, and are the only interlayer tunneling terms kept in the Bistritzer-MacDonald model\cite{BMModel,Balents19}.

As mentioned in the introduction, the anomalous decrease of the bandwidth near the magic twist angle promotes the importance of the next-to-leading order terms in setting the anisotropies, thus selecting from the nearly degenerate manifold of correlated states that are obtained if only the leading order terms are kept. Instead of listing all of the consequences of the above symmetries on such higher order terms, here we only mention in passing that $C_2\mathcal{T}$ and the combined operation $C_{2x}\mathcal{R}_y$ will be seen to allow for a particularly interesting inter-layer tunneling contact term which, as shown in the companion paper, is the main source of the particle-hole symmetry breaking in the model of Ref.~\cite{KaxirasPRB16}, but which is altogether absent in the Slater-Koster type models\cite{MagaudNL10,KoshinoPRB12,KangVafekPRX}.

\acknowledgments
O.~V.~is supported by NSF DMR-1916958 and is partially funded by the Gordon and Betty Moore Foundation's EPiQS Initiative Grant GBMF11070, National High Magnetic Field Laboratory through NSF Grant No.~DMR-1644479 and the State of Florida. J.~K.~acknowledges the support from the NSFC Grant No.~12074276, the Double First-Class Initiative Fund of ShanghaiTech University, and the start-up grant of ShanghaiTech University. Part of this work was performed at the Aspen Center for Physics, which is supported by National Science Foundation grant PHY-1607611.

\appendix

\section{Quadratic Order}\label{App:quadratic}
In Sec.~\ref{Sec:DeriveEffModel}, we expand the effective continuum model to the first order of $\fvec y = \fvec X^{\parallel} - \fvec X'^{\parallel}$. Numerically, we found that the intra-layer part needs to be expanded to the second order of $\fvec y$ to achieve the agreement between the two dispersion produced by $H_{eff}^{K}$ and the microscopic tight binding model $H_{tb}$. To the second order of $\fvec y$, we have
\begin{align}
	& \Psi_{j,S}^{\dagger}(\fvec x + \frac{\fvec y}2) \Psi_{j',S'}(\fvec x - \frac{\fvec y}2) 	\simeq   \Psi_{j,S}^{\dagger}(\fvec x) \Psi_{j',S'}(\fvec x)+ \nonumber \\
	&  \frac{\fvec y}2 \cdot \left[ (\fvec \nabla \Psi_{j,S}^{\dagger}(\fvec x)) \Psi_{j', S'}(\fvec x) - \Psi_{j,S}^{\dagger}(\fvec x) (\fvec \nabla \Psi_{j', S'}(\fvec x)) \right]+ \nonumber \\
	& \frac18 \fvec y^{\mu} \fvec y^{\nu} \left[ (\partial_{\mu} \partial_{\nu} \Psi^{\dagger}_{j,S}(\fvec x) ) \Psi_{j',S'}(\fvec x) - 2 (\partial_{\mu} \Psi^{\dagger}_{j,S}(\fvec x) ) ( \partial_{\nu} \Psi_{j',S'}(\fvec x)  ) \right. \nonumber \\
	& \left. +  \Psi^{\dagger}_{j,S}(\fvec x) (\partial_{\mu} \partial_{\nu} \Psi_{j',S'}(\fvec x))  \right].
\end{align}

\section{Symmetry}
\label{App:symmetry}
	For a smooth but otherwise arbitrary $\fvec U^{\parallel,\perp}_{j,S}(\fvec x)$ the effective Hamiltonian $H_{eff}^\bK$ in Eq.(\ref{Eqn:EffContH}) is invariant under $C_2\mathcal{T}$ if we simultaneously transform the fermion operators {\it and} the atomic displacement fields\cite{Balents19} as
	\begin{align}
		C_2\mathcal{T}:\;\; & \Psi_{j, S}(\fvec x)  \longrightarrow \Psi_{j, \bar{S}}(- \fvec x), \\
		& \fvec U^{\parallel}_{j,S}(\fvec x) \longrightarrow - \fvec U^{\parallel}_{j, \bar{S}}(-\fvec x), \\
		& \fvec U^{\perp}_{j,S}(\fvec x) \longrightarrow \fvec U^{\perp}_{j, \bar{S}}(-\fvec x),  \label{Eqn:C2TSym}
	\end{align}
	where $\bar{S}$ is the sublattice index different from $S$, provided the microscopic hopping function $t(\fvec y + \fvec U_{j,S}^{\perp} - \fvec U_{j',S'}^{\perp}) = t^*(- \fvec y+ \fvec U_{j,S}^{\perp} - \fvec U_{j',S'}^{\perp})$. This is certainly satisfied for the Slater-Koster type model (\ref{Eqn:SlaterKoster}); $C_2\mathcal{T}$ is also satisfied for the orientation dependent hopping function of Ref.\cite{KaxirasPRB16}.
	
	The consequences of this symmetry for the contact term of the inter-layer tunneling part of $H_{eff}^\bK$ (i.e. the first term in the third line of Eq.(\ref{Eqn:EffContH})) can be seen if we assume that the $\fvec U^{\parallel,\perp}_{j,S}(\fvec x)$ is independent of $S$. Then the said term can be expressed as
	\begin{align}
		& \sum_{\fvec G} \int\rmd^2\fvec x\
		e^{i(\fvec G + \fvec K)\cdot \left(\fvec U^\parallel_{t}(\fvec x) - \fvec U^\parallel_{b}(\fvec x) \right)}
		\Psi^{\dagger}_{t,S}(\fvec x)\Psi_{b,S'}(\fvec x)\times\nonumber\\
		&
		T^{\fvec G}_{SS'}\left( \nabla_{\fvec x}\fvec U^\parallel_{t}(\fvec x),\nabla_{\fvec x}\fvec U^\parallel_{b}(\fvec x),\fvec U^\perp_{t}\left(\fvec x\right) - \fvec U^\perp_{b}(\fvec x)\right)  + h.c.
	\end{align}
	Note that Eqs. (\ref{Eqn:tHermitian}-\ref{Eqn:tTRS}) guarantee that $H^{\bK}_{eff}$ is Hermitian.
	Now, $C_2\mathcal{T}$ forces
	\begin{eqnarray}
		&&    T^{\fvec G}_{SS'}\left(\nabla_{\fvec x}\fvec U^\parallel_{t}(\fvec x),\nabla_{\fvec x}\fvec U^\parallel_{b}(\fvec x),\fvec U^\perp_{t}\left(\fvec x\right) - \fvec U^\perp_{b}(\fvec x)\right)=\nonumber\\
		&&\sigma^x_{SS_1}T^{\fvec G *}_{S_1S_2}\left(\nabla_{\fvec x}\fvec U^\parallel_{t}(\fvec x),\nabla_{\fvec x}\fvec U^\parallel_{b}(\fvec x),\fvec U^\perp_{t}\left(\fvec x\right) - \fvec U^\perp_{b}(\fvec x)\right)
		\sigma^x_{S_2S'}.\nonumber\\
	\end{eqnarray}
	This implies that (temporarily suppressing its arguments),
	\begin{equation}
		T^{\fvec G}_{SS'}=
		1_{SS'}W_0^{\fvec G}+\sigma^x_{SS'}W_1^{\fvec G}+\sigma^y_{SS'}W_2^{\fvec G}+i\sigma^z_{SS'}W_3^{\fvec G},
	\end{equation}
	where $W_j^{\fvec G}\left(\nabla_{\fvec x}\fvec U^\parallel_{t}(\fvec x),\nabla_{\fvec x}\fvec U^\parallel_{b}(\fvec x),\fvec U^\perp_{t}\left(\fvec x\right) - \fvec U^\perp_{b}(\fvec x)\right)$ are purely real functions.
	
	Another useful constraint can be obtained from the combination of $C_{2x}$ and $\mathcal{R}_y$
	\begin{align}
		C_{2x}\mathcal{R}_y:\;\; & \Psi_{j, S}(\fvec x)  \longrightarrow \Psi_{\bar{j}, S}( \fvec x),  \\
		& \fvec U^{\parallel}_{j,S}(\fvec x) \longrightarrow  \fvec U^{\parallel}_{\bar{j}, S}(\fvec x), \\
		& \fvec U^{\perp}_{j,S}(\fvec x) \longrightarrow -\fvec U^{\perp}_{\bar{j}, S}(\fvec x), \label{Eqn:RyC2xSym}
	\end{align}
	under which $H_{eff}^\bK$  in Eq.(\ref{Eqn:EffContH}) is also invariant if $t(\fvec y + \fvec U^{\perp}_{j,S} - \fvec U^{\perp}_{j',S'}) = t(-\fvec y + \fvec U^{\perp}_{j,S} - \fvec U^{\perp}_{j',S'})$ as satisfied by the Slater-Koster type model (\ref{Eqn:SlaterKoster}); $C_{2x}\mathcal{R}_y$ is also a symmetry of the orientation dependent hopping function of Ref.\cite{KaxirasPRB16}.
	Then, $C_{2x}\mathcal{R}_y$ forces
	\begin{eqnarray}
		&&    T^{\fvec G}_{SS'}\left(\nabla_{\fvec x}\fvec U^\parallel_{t}(\fvec x),\nabla_{\fvec x}\fvec U^\parallel_{b}(\fvec x),\fvec U^\perp_{t}\left(\fvec x\right) - \fvec U^\perp_{b}(\fvec x)\right)=\nonumber\\
		&&T^{\fvec G *}_{S'S}\left(\nabla_{\fvec x}\fvec U^\parallel_{b}(\fvec x),\nabla_{\fvec x}\fvec U^\parallel_{t}(\fvec x),\fvec U^\perp_{t}\left(\fvec x\right) - \fvec U^\perp_{b}(\fvec x)\right).
	\end{eqnarray}
	The anti-Hermiticity of $i\sigma^z$ implies that $W^\bG_3$ must be {\it odd} under $\nabla_{\fvec x}\fvec U^\parallel_{t}(\fvec x)\leftrightarrow \nabla_{\fvec x}\fvec U^\parallel_{b}(\fvec x)$. But, the contact inter-layer tunneling term in Eq.(\ref{Eqn:EffContH})
	is clearly {\it even} under
	$\nabla_{\fvec x}\fvec U^\parallel_{t}(\fvec x)\leftrightarrow \nabla_{\fvec x}\fvec U^\parallel_{b}(\fvec x)$.
	Therefore, for the Slater-Koster type models $W^\bG_3=0$. This is indeed what we find from detailed analysis presented in the companion paper.
	On the other hand, for the model based on the ab-initio hopping integrals~\cite{KaxirasPRB16}, there is an additional dependence of the hoppings on the orientation of hopping vector to the nearest neighbor vectors. In this case, $W^\bG_3\neq 0$ even for a rigid twist. As we show in the companion paper, this term gives the largest contribution to the particle-hole asymmetry.

\end{document}